# Low-threshold induced side-scattering instability in the edge transport barrier at O-mode ECRH experiments in magnetic fusion devices

E.Z. Gusakov, A.Yu. Popov

*Ioffe Institute, Saint-Petersburg, Russia*

The lower hybrid wave trapping in a tokamak edge transport barrier is predicted. This effect makes possible excitation of the low-power-threshold absolute parametric decay instability leading to side-scattering of the ordinary microwave pump in the electron cyclotron resonance heating experiments. The instability can result in broadening of the power deposition profile both in present day machines and in ITER.

1. **Introduction**

   Three-wave interactions occurring in nonlinear media for waves satisfying momentum and energy conservation principles (or decay conditions) give rise to so called parametric decay instabilities (PDIs) [1]. These instabilities, leading to the generation of daughter waves in the media, are excited in the presence of a strong pump wave if its amplitude exceeds a certain threshold determined by the daughter waves' energy losses [1]. In inhomogeneous media the PDI threshold is often determined by convective losses of daughter waves from the decay region [2,3]. In plasmas the PDIs cause anomalous absorption and reflection of electromagnetic pump waves if their thresholds are overcome, as in the case of ionosphere modification experiments [4] and laser fusion [5-7]. However, the theoretical analysis [8] predicted extremely high pump power-thresholds for any parametric decay instability that can accompany microwave propagation and damping in magnetic fusion devices in electron cyclotron resonance heating (ECRH) experiments. Therefore, until recently this method was thought to be free from anomalous phenomena and thus provide a predictable pump power deposition. It is widely used in current toroidal magnetic fusion devices and is planned for application in future fusion experiments. In particular, the fundamental harmonic ordinary mode (O1-mode) ECRH system is in preparation both for electron component heating and for controlling the neoclassical tearing mode in ITER and also discussed for DEMO. However, over the last decade a number of anomalous phenomena – anomalous microwave backscattering [9,10], ion acceleration [11,12], an evident broadening of the ECRH power deposition profile [13,14] and radiation at sub-harmonics of the gyrotron frequency [15] – were discovered in ECRH experiments in different toroidal devices. These anomalous effects were observed at the ECRH power level much smaller than the PDI threshold value predicted for inhomogeneous plasma in [8], thus indicating limitations of this theoretical model. Its extension, properly taking into account the presence of a non-monotonic (hollow) density profile often observed in experiments with ECRH [16,17], was proposed and then developed [18-21]. The key mechanism underlying the novel model is related to trapping of one of the daughter waves (or even both of them) in the presence of a non-monotonic density profile. Such localization leads to the complete suppression of energy losses by the daughter wave from the decay layer and, consequently, to a drastic decrease (several orders of magnitude) in the decay instability power-threshold. The new model was used to analyze the most dangerous scenarios of low power-threshold parametric decays of both the second harmonic extraordinary microwave [20,21] and the fundamental harmonic ordinary microwave [22,23] in the vicinity of the local maximum of a non-monotonic density profile, leading to the excitation of localized upper hybrid waves [24]. The possibility of substantial anomalous absorption had also been experimentally demonstrated in the model experiment [25]. These results stimulated experimental activity at ASDEX-Upgrade [26,27] and Wendelstein 7-X [28], where low-threshold PDIs were found in various sets of ECRH experiments. Numerical simulations confirm the parametric excitation of upper hybrid eigenmodes at the nonmonotonic density profile [29]. However, the non-monotonic density profile does not appear to be the only cause of low power-threshold PDIs. In the present paper it is shown that quite unexpectedly the excitation of these phenomena can occur in the steepest region of the density profile, *i.e.* in the edge transport barrier, which usually exists in high performance discharges, and where, at first sight, the convective losses of daughter waves should be maximum and the conclusions of the theoretical analysis [8] should be justified. Nevertheless, it has been shown that the specific transparency domains afforded to intermediate frequency waves in regions with a high density gradient [30,31] result in their easy parametric excitation. As shown below, these waves can be trapped both in the direction of plasma inhomogeneity due to the edge transport barrier and along the magnetic field in its ripples associated with a finite number of toroidal magnetic field coils. Since the power thresholds of parametric decays leading to the excitation of localized daughter waves [18-24] turn out to be much lower than those predicted for the generation of non-localized daughter waves [8], absolute parametric decay instability (PDI) of megawatt microwave beams seems possible in future O1-mode ECRH experiments at ITER. This instability leads to the excitation of anomalously scattered ordinary waves and low hybrid waves (LHWs) trapped in the direction of plasma inhomogeneity and along the magnetic field.

2. **Intermediate frequency wave trapping in strongly inhomogeneous plasmas**

The usual approach to the analysis of intermediate frequency waves in inhomogeneous magnetized plasmas is the WKB approximation, which leads to the same conclusions on the wave transparency regions as the homogeneous plasma theory. However, strong plasma inhomogeneity at the plasma edge combined with a large value of the non-diagonal dielectric tensor component can lead to a significant change in the wave transparency [30,31], creating new transparency regions. The wavelength in this case remains much smaller than the plasma inhomogeneity scale length and therefore the effect can be accounted for in the WKB approximation modified by adding terms proportional to the derivatives of the dielectric tensor components [30]. To illustrate the manifestation of this phenomenon in the tokamak edge transport barrier (ETB), we introduce the local Cartesian coordinate system $(x, y, z)$. The coordinate $x$ is related to the flux surface label, $y$ - the coordinate perpendicular to the magnetic field line on the magnetic surface and $z$ - the coordinate directed along the magnetic field line. The magnetic system of ITER – the installation with a major radius $R = 6.2$ m – is designed with a number of toroidal coils $N = 36$. The magnetic field in a narrow layer in the ETB has the form $B = \bar{B}\left(1 - \delta(x, y)\cos(Nz/R)\right)$ with the magnetic ripples amplitude of $\delta \approx 1\%$. A possibility of reduction of magnetic ripple using ferrite inserts from the value exceeding in the equatorial plane at the outboard plasma edge $\delta \approx 1\%$ down to $\delta \approx 0.3\%$ is discussed [32,33]. The magnetic field at the plasma edge is taken $\bar{B} = 4$ T. The amplitude of the electrostatic low hybrid wave

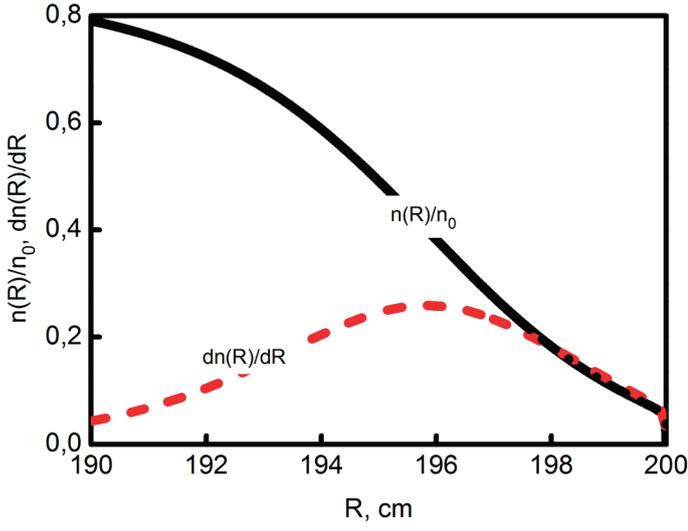 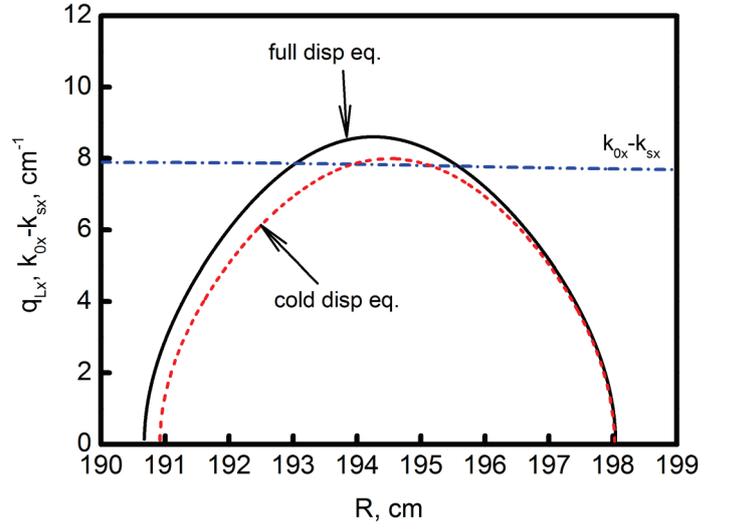

Figure 1. The density profile normalized to the density at the magnetic axis ($n_0 = 1 \times 10^{20}\,\text{m}^{-3}$, thick solid line), and the profile of its derivation (dashed line).

Figure 2. The LHW ($f_L = 1.12\,\text{GHz}$, $q_y^* = 7.99\,\text{cm}^{-1}$) dispersion curves $q_{Lx} = q_{Lx}(q_y^*)$ calculated by solving both the full dispersion equation (solid line), which properly takes into account thermal effects, and the cold dispersion equation (dashed line). The dashed dotted curve shows $k_x(\omega_0) - k_x(\omega_s, q_y^*)$, $f_0 = 170\,\text{GHz}$, $f_s = 168.88\,\text{GHz}$. The ion temperature at the density pedestal is $T_i = 1\,\text{keV}$.

$\phi(\mathbf{r}) = \psi(x,z)\exp(iq_y y + i\omega_L t)/2 + c.c.$ at the $\omega_L$ frequency is described by the Poisson equation
$\hat{D}_{LHW}\psi = \left(\varepsilon(\omega_L)\Delta_\perp + \partial_x\varepsilon(\omega_L)\partial_x + \partial_x g(\omega_L)q_y + \eta(\omega_L)\partial_{zz}\right)\psi = 0$ where $\varepsilon$, $g$, $\eta$ are the components of the cold plasma dielectric tensor [34], $\Delta_\perp = \partial_{xx} - q_y^2$, $\partial_\zeta = \partial/\partial\zeta$, $\zeta = x,y$ and the term $\sim \partial_x \varepsilon(\omega_L)$ being much smaller than the term $\sim \partial_x g(\omega_L)$ by a factor of $\omega_L/\omega_{ce} \ll 1$ will be further neglected. The term $Q = \partial_x g/\varepsilon\big|_{\omega_L}$ in the operator $\hat{D}_{LHW}$ depends on the coordinates $x$ and $z$. To illustrate the dependence on $x$, we take a density profile close to that expected in ITER in the edge transport barrier [35] (see figure 1, solid line). The same figure shows its spatial derivative (dashed line), which has a local maximum at $x_m$ in the barrier region. Since the function $Q$ depends on the magnetic field, it has also a local minimum along the toroidal direction at $z = 0$ between the two adjacent toroidal magnetic field coils where the pump power is launched. We approximate $Q$ by a quadratic dependence on both coordinates $Q \approx Q_0(1 - (x - x_m)^2/(2l_x^2) + z^2/(2l_z^2))$ around $x = x_m$, $z = 0$, which is strictly speaking valid only in the vicinity of the magnetic field minimum. Using this expansion yields

$\hat{D}_{LHW}\psi = \varepsilon(\omega_L, x_m)\left(\partial_{xx} + Q_0 q_y - q_y^2 - K_x^4(x - x_m)^2\right)\psi - |\eta(\omega_L, x_m)|\left(\partial_{zz} - K_z^4 z^2\right)\psi = 0$ (1)

where $K_x = (Q_0 q_y/(2l_x^2))^{1/4}$ and $K_z = K_x(l_z^2|\eta(\omega_L)|/l_x^2)^{-1/4}$. The solution to equation (1), representing the

LHW trapped both along the magnetic field and in the radial direction

$$\psi(\mathbf{r}) = \psi_{p,r} f_p(K_x(x-x_m)) f_r(K_z z), \quad \psi_{p,r} = const, \qquad (2)$$

is expressed in terms of the Hermite polynomials $f_p(Kx) = \sqrt{K/(\sqrt{\pi}2^p p!)} \exp(-K^2 x^2/2) H_p(Kx)$. This expression is valid in the case the wave localization region is much smaller than the distance between the magnetic coils. In the opposite case the WKB solutions of (1) should be used. Substitution of (2) into (1) gives the dispersion relation, which defines the quantization condition for the mode eigenfrequency

$$D_{LHW}(\omega_L^{p,r}) = \varepsilon(\omega_L^{p,r})(Q_0(\omega_L^{p,r})q_y - q_y^2 - (2p+1)K_x^2(\omega_L^{p,r})) + (2r+1)|\eta(\omega_L^{p,r})|K_z^2(\omega_L^{p,r}) = 0. \qquad (3)$$

These trapped LHWs, which propagate almost across the magnetic field, exist only in strongly inhomogeneous plasmas, where there are regions of transparency for those with a positive poloidal number. If the density gradient is small or $q_y^2$ is too large, the plasma for such LHWs turns out to be evanescent. It should be noted that the trapped LHW has another noteworthy property. According to (3) its group velocity along $y$ determined as $v_{gy} = \partial D_{LHW}/\partial q_y / \partial D_{LHW}/\partial \omega_L^{p,r}\big|_{\omega_L^{p,r}, q_y}$ takes the zero value at $q_y^* = Q_0/2$.

### 3. Low-threshold parametric excitation of the LHW trapped in the ETB

We will show that the 2D localized LHWs can be easily excited in O-mode ECRH experiments. Given the geometry of future experiments in ITER, we consider an ordinary pump wave propagating perpendicular to the magnetic field along the $x$ - coordinate to the plasma core with its polarization vector being directed mostly along the magnetic field. By means of the WKB approximation it is represented as

$\mathbf{E}_0 = \mathbf{e}_z A_0 \exp\left(i\int_0^x k_x(\omega_0, x')dx' - i\omega_0 t\right) + c.c.$  where  $A_0 = \sqrt{2P_0/(cw^2)}\, n_x(\omega_0, x)^{-1/2} \exp(-(y^2+z^2)/2w^2)$,

$P_0$ - the pump power, $w$ - the width of a beam (in what follows we assume $w = 2$ cm), $c.c.$ - the term derived from the first one by complex conjugation, $n_{0x} = ck_x(\omega_0)/\omega_0 = \sqrt{1-\omega_{pe}^2/\omega_0^2}$ - the wave refraction index, $\omega_{pe}$ - the electron plasma frequency. We analyse the pump wave decay in the ETB into the trapped LHW (2) and the ordinary wave $\mathbf{E}_s(\mathbf{r}) = \mathbf{e}_z A_s(x)\exp(iq_y y + i\omega_s t)/2 + c.c.$, whose amplitudes are described by the set of nonlinearly coupled equations

$$\begin{cases} \hat{D}_s A_s = \Delta_\perp A_s + \omega_s^2/c^2 \eta(\omega_s) A_s = -i\kappa_{nl}\omega_s/c \Delta_\perp A_0^* \psi \exp\left(-i\int_0^x k_{0x}(x')dx'\right) \\ \hat{D}_{EPW}\psi = i\kappa_{nl}c/\omega_s \Delta_\perp E_0 A_s \exp\left(i\int_0^x k_{0x}(x')dx'\right) \end{cases} \qquad (4)$$

where $\kappa_{nl} = \omega_{pe}^2/(2\omega_0\omega_{ce}\overline{B})$ - the nonlinear coupling coefficient [34], $\omega_{ce}$ - the electron cyclotron frequency. To illustrate the case we plot in figure 2 the LHW ($f_L = 1.12$ GHz, $q_y^* = 7.99$ cm$^{-1}$) dispersion curves calculated by solving both the full dispersion equation (solid line), which properly takes into account ion thermal effects, and the cold dispersion equation (dashed line). It can be seen that the LHW is localised in the ETB and the cold equation describes it with reasonable accuracy. The dashed dotted curve shows

$k_x(\omega_0) - k_x(\omega_s, q_y^*)$, $f_0 = 170$ GHz, $f_s = 168.88$ GHz. The ion temperature at the density pedestal is taken to be $T_i = 1$ keV. At the points where the dashed and dashed dotted curves intersect, the decay condition $\Delta K = q_{Lx}(\omega_L, q_y^*) - k_x(\omega_0) + k_x(\omega_s, q_y^*) = 0$ is fulfilled and the decay instability becomes possible. Solving the first equation in the set of equations (4), we obtain $A_s = -i\kappa_{nl}(\omega_s/c)G_s\{\Delta_\perp A_0 \psi\}$ where $G_s\{...\} = \frac{ic}{2\omega_s}\int_{-\infty}^x \frac{dx'\{...\}}{\sqrt{n_{sx}(x)n_{sx}(x')}} \exp\left(-i\int_x^{x'}(k_{0x}(x'') - k_{sx}(q_\zeta^*, x''))dx''\right)$ is a part of the Green's function of this equation, in which the dominant term describing the resonance interaction of waves is held. Substituting $A_s$ into the RHS of the second equation of system (4), we get the equation for the LHW potential

$$\hat{D}_{LHW}\psi = \kappa_{nl}^2 \Delta_\perp \left( A_0 G_s \{\Delta_\perp (A_0^* \psi)\} \right) \tag{5}$$

To find a solution to (5), we utilize the procedure of perturbation theory [36]. In its first step, we neglect the nonlinear pumping described by the RHS of equation (5) and the daughter LHW energy losses along the $y$ direction. A solution to the homogeneous version of equation (5) is determined through equation (2). In the second step of the perturbation procedure, we take into account the LHW energy loss along $y$. For such a wave with the poloidal number $q_y^* = Q_0/2$ the group velocity tends to zero, and therefore the only mechanism of energy loss from the decay region in the $y$ direction is diffraction, which is a slower process than convection. Therefore the LHWs, possessing the poloidal wavenumber close to this value are most unstable and excited first of all during the pump wave decay. The nonlinear interaction and energy loss make the amplitude $\psi_{p,r}$ no more constant, i.e. $\psi_{p,r} \to \psi_{p,r}(t,y)$ with $p = 12$, $r = 2$ at $\delta \simeq 0.75\%$ and $p = 9$, $r = 2$ at $\delta \simeq 0.3\%$. Substituting (2) into (5), multiplying both sides of the latter by $f_p(K_x x)^* f_r(K_z z)^*$ and integrating over $x$ and $z$, we arrive at the equation describing the absolute PDI

$$\left(\partial/\partial t + i\Lambda_y \partial^2/\partial y^2\right)\psi_{p,r} = \gamma_0 \exp(-y^2/w^2)\psi_{p,r}, \tag{6}$$

$\gamma_0 = i\frac{\omega_{pe}^4 (k_x^2(\omega_0) + q_y^{*2})^2}{\omega_0^2 \omega_{ce}^2 |<\partial_{\omega_L} D_{LHW}>|} \frac{2P_0}{cn_{0x}w^2 \overline{B}^2} \left| \int_{x_m}^\infty dz |f_r(z)|^2 \exp\left(-\frac{z^2}{w^2}\right) \int_{-\infty}^\infty dx f_p(x)^* G_s \left\{ \exp\left(i\int_{x'}^x k_x(\omega_0)dx''\right) f_p(x') \right\} \right|$ is

the pumping rate, $\Lambda_y = \varepsilon(\omega_L, x_m)/<\partial_{\omega_L} D_{LHW}>$ - the diffraction coefficient and $<...>$ - the averaging over the LHW localization region. Equation (6) describes the LHW exponential growth, which occurs when the pump power exceeds the threshold value $P_0^{th}$. If the pump power is significantly higher than the threshold value, $P_0 \gg P_0^{th}$, we can expand the function $\exp(-y^2/w^2) \approx 1 - y^2/w^2$ and get an analytical approximation for the exponentially growing solution [20]

$$\psi_{p,r}(t,y) = \exp(\gamma_{ins}^s t + i\delta\omega_{ins}^s t) f_s(y/\delta_y), \quad \delta_y = \Lambda_y^{1/4} w^{1/2}/\sqrt[4]{\gamma_0} - \exp(-i\pi/8 - i\arg(\gamma_0/4)), \tag{7}$$

where

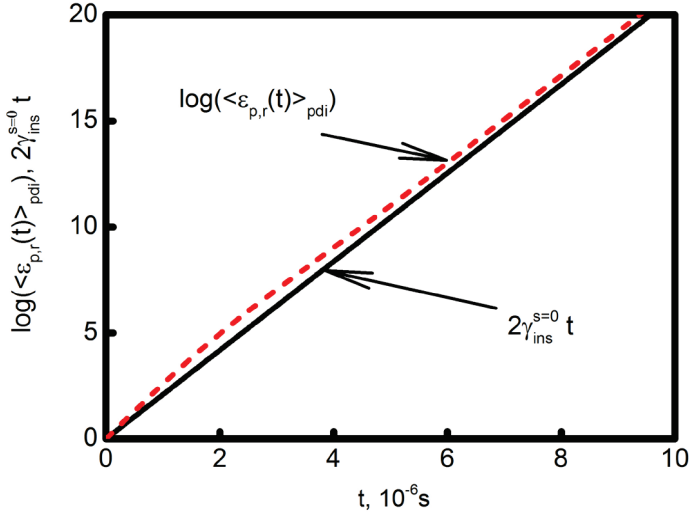
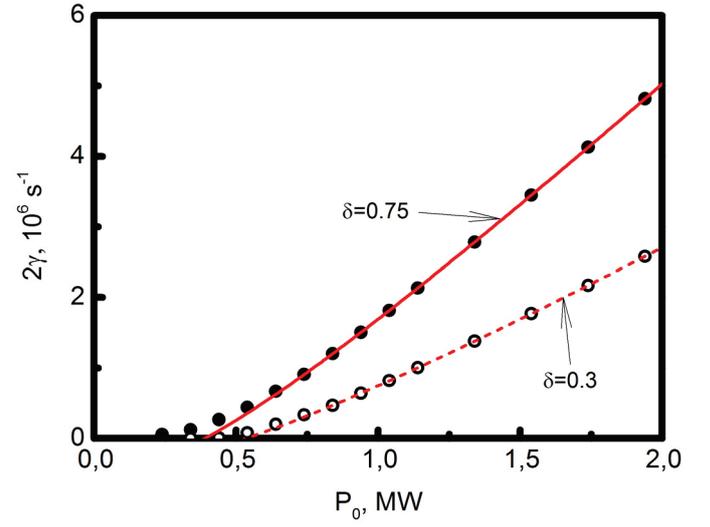

Figure 3. The amplification coefficients calculated numerically (dotted curve) and predicted analytically by equation (8) (solid curve) for the most unstable fundamental mode $s = 0$ and $P_0 = 1$ MW.

Figure 4. Dependence of the growth rate on the pump power, $w = 2$ cm. The solid ($\delta \simeq 0.75\%$) and dashed ($\delta \simeq 0.3\%$) curves are given by eq. (8). The scattered circles (closed - $\delta \simeq 0.75\%$, open - $\delta \simeq 0.3\%$) are numerical solutions of eq. (4).

$$\gamma_{ins}^s = \gamma_0' - (2s+1)\sqrt{|\gamma_0|\Lambda_y / w^2} \sin\left(\arctan(\gamma_0''/\gamma_0')/2 + \pi/4\right)$$
$$\delta\omega_{ins}^s = \gamma_0'' + (2s+1)\sqrt{|\gamma_0|\Lambda_y / w^2} \cos\left(\arctan(\gamma_0''/\gamma_0')/2 + \pi/4\right) \qquad (8)$$

with $s = 0, 1...$ Though equation (8) is no more valid when the pump wave with the power $P_0 \geq P_0^{th}$ is marginally unstable, we can use it to estimate roughly the PDI power-threshold, which is determined by the LHW diffractive loss. To this end we set $\gamma_{ins}^0 = 0$ in equation (8), which gives the condition for $P_0^{th}$

$$\gamma_0'(P_0^{th}) = (2s+1)\sqrt{|\gamma_0(P_0^{th})|\Lambda_y / w^2} \sin\left(\arctan(\gamma_0''(P_0^{th})/\gamma_0'(P_0^{th}))/2 + \pi/4\right) \qquad (9)$$

Then, we solve (6) numerically and plot the results in figure 3, where the temporal dependence of the wave energy in the region of the pump beam localization $<\varepsilon(t)>_{pdi} = T_e / (\sqrt{\pi} w) \int_{-\infty}^{\infty} dy \exp(-y^2/w^2) |\psi_{p,r}(t,y)|^2$ is shown in semi-logarithmic scale. The solid curve shows the same dependence but analytically predicted by (8). Being close to each other, they indicate a temporal growth of the LHW amplitude, which confirms the excitation of absolute PDI. A reasonable agreement between the numerical and analytical dependences is demonstrated. Figure 4 shows the dependence of the instability growth rate defined by equation (8) on the pump power at the pump beam width of 2 cm and magnetic field ripples $\delta \simeq 0.3\%$ (dashed curve) and $\delta \simeq 0.75\%$ (solid curve). The circles (open - $\delta \simeq 0.3\%$ and closed - $\delta \simeq 0.75\%$) are the numerical solutions to equation (6). The numerically calculated power thresholds are $P_0^{th} = 419$ kW and $P_0^{th} = 256$ kW, correspondingly. Rough analytical estimates of the instability power thresholds in these cases provided by equation (9) overestimate their real values. It should be stressed that the obtained values of the absolute instability power threshold is two orders of magnitude smaller than the value predicted for the induced scattering instability by the standard theory [8], thus making this absolute PDI inevitable in ITER and leading to the risk of strong anomalous side-scattering of the ECRH power similar to that observed in laser fusion experiments [37]. At a pump power much higher than the power threshold of the instability excitation, analytical dependence (8) describes the growth rate with good accuracy. For the expected value of the ECRH power in a single beam of 1 MW planned for ITER the growth rate is equal to $1\times10^6$ s$^{-1}$. However, in the case when several heating beams cross the LHW localisation region, the growth rate will increase proportionally to the number of beams. It should be mentioned that if the microwave beams are launched not in the magnetic field minimum region, which is more complicated technically, and are not crossing the LHW localisation region the instability threshold increases. Knowing the values of $q_y^*$ and $k_x(\omega_s, q_y^*)$, it is also possible to predict the angle at which the pump wave scatters anomalously and using the ray tracing procedure to determine a localisation of the scattered wave absorption. Depending on the saturation level the side-scattering PDI can lead to anomalous broadening of the ECRH power deposition profile. It should be noted that the alternative explanation of the broadening of the power deposition profile was proposed in [38] based on effect of the pump wave scattering off turbulent drift-wave density fluctuations at the plasma edge [39-42]. The effect of low-threshold induced side-scattering PDI predicted in the present paper for the ITER O-mode ECRH experiment can be investigated experimentally at contemporary tokamaks possessing edge transport barrier and utilizing O-mode ECRH, in particular, at ASDEX-Upgrade. The O2-ECRH experiment planned at ASDEX-Upgrade [43] is especially suitable for that because of modest single-pass absorption allowing measurements of the side-scattering signal. According to the estimation based on the model developed in the present paper for the ASDEX-Upgrade ETB parameters [44], the decay threshold is of the order of 400 kW with magnetic ripple amplitude of 1% and a beam width of 2 cm.

4.  **Conclusions**

It is predicted that the electron plasma wave trapping in the edge transport barrier leads to the low power-threshold induced side-scattering absolute parametric decay instability of ordinary microwaves. At the magnetic field ripple $\delta \simeq 0.75\%$, the minimum power-threshold of the PDI leading to scattering at the angle of $0.1\pi$ with frequency down-shift of 1.1 GHz is as small as 256 kW in a single microwave beam. At magnetic ripple amplitude of $\delta \simeq 0.3\%$, the minimum power-threshold of the PDI leading to scattering at the angle of $0.12\pi$ with frequency down-shift of 1.12 GHz is equal to 419 kW. The obtained values of the power-threshold of absolute instability are two orders of magnitude smaller than the value predicted for the induced scattering instability by the standard theory [8]. This nonlinear effect, leading to anomalous reflection of heating power, could easily occur in O1-mode ECRH experiments at ITER, where multiple megawatt pump beams are planned for utilization. Undoubtedly, this effect can have a significant impact on the performance of the ECRH system at ITER and should be taken into account seriously when planning the future experiments. The theoretical model developed in the present paper and the role of the side-scattering PDI in the broadening of the ECRH power deposition profile can be checked in the ASDEX-Upgrade in O2 – mode experiment.

**Acknowledgements.** *The PDI analytical treatment is supported under the Ioffe Institute state contract 0040-2019-0023, whereas the numerical modelling is supported under the Ioffe Institute state contract 0034-2021-0003*